\documentclass{aa}  
\pdfoutput=1 % For arxiv
\usepackage{graphicx}
\usepackage{txfonts}
\usepackage{comment}
\usepackage{xcolor}
\newcommand{\HI}{H\,\textsc{i} }
\newcommand{\hi}{H\,\textsc{i}}
\newcommand{\apx}{$\sim$ }
\newcommand{\pc}{$\%$ }

\newcommand{\target}{B2~0258+35 }
\newcommand{\targets}{B2~0258+35}
\newcommand{\eg}[1]{\citep[e.g.][]{#1}}
\newcommand{\kmps}{km~s$^{-1}$ }
\newcommand{\kmpss}{km~s$^{-1}$}
\newcommand{\p}[1]{$^{-#1}$}
\newcommand{\pp}[1]{$^{#1}$}

\newcommand{\kmpsc}{km~s$^{-1}$~}

\begin{document}

   \title{Feedback from low-luminosity radio galaxies: B2 0258+35}

   \author{Suma Murthy\inst{1,2},
           Raffaella Morganti\inst{2,1},
          Tom Oosterloo\inst{2,1},
          Robert Schulz\inst{2},
          Dipanjan Mukherjee\inst{3},
          Alexander Y. Wagner\inst{4},
          Geoffrey Bicknell\inst{5},
          Isabella Prandoni\inst{6},
          Aleksandar Shulevski\inst{7}}

   \institute{Kapteyn Astronomical Insititute, University of Groningen, Postbus 800, 9700 AV Groningen, The Netherlands
              \email{murthy@astro.rug.nl}
         \and
             ASTRON, the Netherlands Institute for Radio Astronomy, Postbus 2, 7990 AA, Dwingeloo, The Netherlands
         \and         
             Dipartimento di Fisica Generale, Universita degli Studi di Torino , Via Pietro Giuria 1, 10125 Torino, Italy
         \and
             Center for Computational Sciences, University of Tsukuba, 1-1-1 Tennodai, Tsukuba, Ibaraki, 305-8577
         \and 
             Australian National University, Research School of Astronomy and Astrophysics, Cotter Rd., Weston, ACT 2611, Australia
          \and
             INAF - Istituto di Radioastronomia, via P. Gobetti 101, 40129 Bologna, Italy
           \and
             Anton Pannekoek Institute for Astronomy, University of Amsterdam, Postbus 94249, 1090 GE Amsterdam, The Netherlands}

   \date{Received ..., accepted ...}

% \abstract{}{}{}{}{} 
% 5 {} token are mandatory
 
  \abstract{Low-luminosity radio-loud active galactic nuclei (AGN) are of importance in studies concerning feedback from radio AGN since a dominant fraction of AGN belong to this class. We  report high-resolution Very Large Array (VLA) and European VLBI Network (EVN) observations of \HI 21cm absorption from a young, compact steep-spectrum radio source, B2~0258+35, nested in the early-type galaxy NGC~1167, which contains a 160 kpc \HI disc.  Our VLA and EVN \hi\ absorption observations, modelling, and comparison with molecular gas data suggest that the cold gas in the centre of NGC 1167 is very turbulent (with a velocity dispersion of \apx~90~\kmpsc) and that this turbulence is induced by the interaction of the jets with the interstellar medium (ISM). Furthermore, the ionised gas in the galaxy shows evidence of shock heating at a few kpc  from the radio source. These findings support the results from numerical simulations of radio jets expanding into a clumpy gas disc, which predict that the radio jets in this case percolate through the gas disc and drive shocks into the ISM at distances much larger than their physical extent. These results expand the number of low-luminosity radio sources found to impact the surrounding medium, thereby highlighting the possible relevance of these AGN for feedback.}


\titlerunning{Low-luminosity AGN: B2 0258+35}
\authorrunning{Murthy et al.}
\maketitle

%-------------------------------------------------------------------

\section{Introduction} \label{introduction}

Active galactic nuclei (AGN) have long been of importance in studies of
galaxy evolution. The energy emitted by an active supermassive black
hole (SMBH) is believed to affect its host galaxy in a number of
different ways, collectively referred to as AGN feedback. These effects
can range from regulating the growth of the SMBH (which  can  explain the
relation between the SMBH mass and the velocity dispersion of the
galactic bulge; e.g. \citealt{Silk98}), and the possible quenching of the star
formation via massive gas outflows \citep[][and references
therein]{Harrison18},  to preventing the cooling of the gas in hot halos, which in
turn can prevent massive galaxies from forming \citep[e.g.][and references therein]{Croton06, McNamara12}.

It is thought that radio-loud AGN     play an important
role mostly in the last:  by preventing the cooling of the
circumgalactic and  intergalactic gas through their large-scale radio
lobes,  they prevent the accretion of gas onto the galaxy. This
effect has been clearly observed in a number of relatively bright radio
AGN that have been studied in X-rays \eg{McNamara09,Randall11,Cavagnolo11, Fabian12}.

However, over the past years evidence has been building up to show
that radio AGN also affect the host galaxy directly through radio
jet--ISM interactions \eg{Oosterloo17, Rodriguez-Adrila17,
Fabbiano18b, Maksym18, Wang11, Finlez18, Alatalo15a, Croston08}.
Low-luminosity radio AGN (L$_{\textrm{1.4 GHz}} \lesssim 10^{24.5} \textrm{W Hz}^{-1}$) are of interest to studies concerning this mechanical mode of feedback since they greatly outnumber their high-luminosity counterparts \citep{Best05}.

Recent numerical simulations \eg{Cielo18, Mukherjee18a} also support the
importance of this mode of feedback from radio AGN. They predict that
jets expanding into a clumpy gas disc affect  the gas kinematics
significantly. The extent of this impact depends on the orientation of the jets with respect to the gas disc, among other
parameters.
If the jets expand directly into the gas disc, then the simulations
predict that they will induce fast outflows of multi-phase gas as well as
strong turbulence. These simulations further predict that the jet--ISM
interaction will generate expanding cocoons that shock the gas in the
disc and also affect the star formation rate in the regions surrounding
the AGN activity. The gas may also, in turn, affect the evolution of the
radio source itself \citep[e.g. morphology and age;][]{Mukherjee18b}.

This mechanical feedback is best studied by probing the
kinematics of different phases of gas affected by the radio jets. Such
detailed studies of radio jet--ISM  interactions can provide constraints to
quantify the AGN feedback effects, which can be incorporated into the
simulations of galaxy evolution. In order to gain a better understanding of how this actually happens, we need to expand the number of cases for which detailed,
high-resolution observations of multiple phases of the gas are available
so that a comparison with the predictions of numerical simulations is
possible.

We  present here one such study of a nearby \citep[\textit{z} = 0.0165;][]{Wegner93}, low-luminosity \citep[L$_{1.4 \rm {GHz}}$ = 2.1 $\times$ 10$^{23}$ W Hz$^{-1}$;][]{Shulevski12} radio source: B2 0258+35. This source is nested in a gas-rich galaxy, NGC 1167, and appears to be interacting strongly with the ISM. The galactic \HI is well studied \citep{Struve10c} and so are the properties of the radio source \citep{Sanghera95, Giroletti05, Brienza18}. This radio galaxy is thus a good candidate to understand the nature of jet--ISM interactions and the possible impact the radio source has on the ambient medium.

\begin{figure}
    \includegraphics[width=\linewidth]{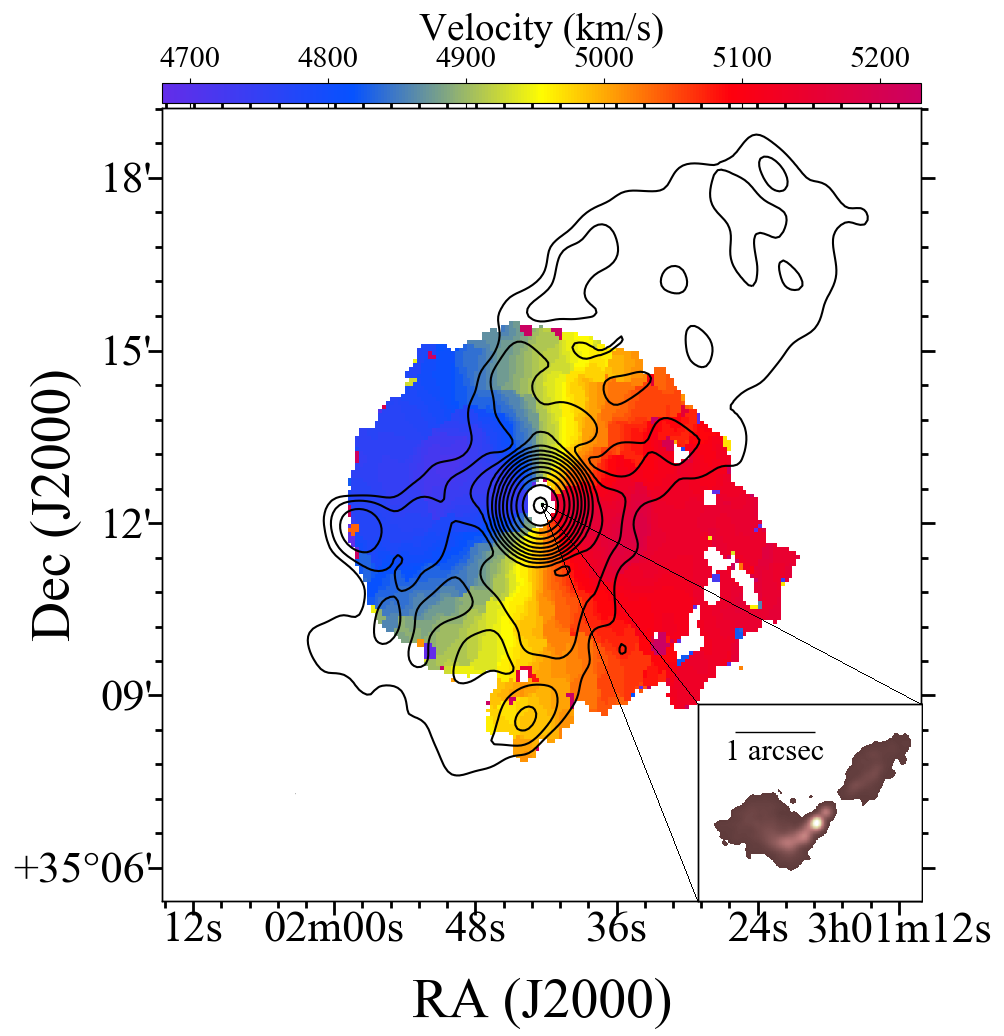}
    \caption{Large-scale \HI disc seen in emission along with the velocity field (in colour). The eastern portion of the rotating disc is the approaching side. \HI is seen in absorption within a radius of \apx 10 kpc \citep[angular resolution of the WSRT observations \apx 29$''$;][]{Struve10c}. The large-scale low surface brightness radio lobes \citep[WSRT observations with a beam of 39$''$ $\times$ 33$''$;][]{Shulevski12} are shown as black contours. The contours start at 400$\mu$Jy beam\p{1} and increase by a factor of 2. The central CSS source \citep[at 22 GHz with VLA A-array at an angular resolution of 0.12$''$][]{Giroletti05} is shown in the inset. The position angle of the \HI disc is 75\pp{\circ} \citep{Struve10c}, while that of the CSS source is 132\pp{\circ} \citep{Giroletti05}.}
    \label{fig:vel_field}
\end{figure}

\target is classified as a compact steep-spectrum (CSS) source \citep{Sanghera95}. The sources in this class are compact with their synchrotron spectrum peaking at a few hundred MHz and are generally considered to be young \eg{Odea98}. Earlier radio continuum studies of \target \citep{Giroletti05} have found the radio source to be about 1 kpc in size, consisting of a core and two jets (see Figure \ref{fig:vel_field}, inset). The northern jet is comparatively faint and the southern jet is bent sharply. This asymmetry both in brightness and morphology suggests an interaction with the ISM, similar to that seen in IC 5063 \citep{Morganti98}. The age of \target is estimated to be between 0.4 Myr \citep{Brienza18} and 0.9 Myr \citep{Giroletti05}. Following the correlation between the age and the size of the source \citep{Orienti14, Murgia03}, it has also been found that the source is smaller than   expected for this age, given an unconstrained expansion. This suggests that it may have been confined within the host galaxy for a protracted length of time \citep{Brienza18}. However, we   note that a hypothesis  based on the above-mentioned correlations alone is quite uncertain, but as we show  in the subsequent sections, the \HI absorption and CO emission studies further strengthen this hypothesis. 

The host galaxy, NGC 1167, was further found to host large-scale (\apx 240 kpc) extremely low surface brightness radio lobes \citep{Shulevski12}, about \apx 110 Myr old \citep{Brienza18}, which make up only 3\pc of the total source luminosity (4.75 mJy arcmin$^{-2}$ at 145 MHz). Though such low surface brightness may imply that these structures are remnants of a previous activity, their spectral indices have been found to be `normal' and not ultra-steep as expected for these sources \citep{Brienza18}. Various possibilities have been put forth to explain this feature, including in situ particle reacceleration, multiple duty cycles, dense ISM smothering the large-scale jets that are still being fuelled at a low level, and magnetic draping enhancing the mixing within the lobes while suppressing the same between the radio lobes and the surrounding medium \citep{Brienza18, Adebahr18}.

NGC 1167 is a gas-rich (M$_{\rm\HI}$ = 1.5 $\times$ 10$^{10}$ M$_\odot$) early-type galaxy, with a regularly rotating 160 kpc \HI disc (seen in emission; Figure~\ref{fig:vel_field}). The disc has very regular kinematics within a radius of 65 kpc, and only in the very outer parts does it show signs of interactions, likely with a satellite galaxy. This indicates that the galaxy has not undergone a major merger in the last few billion years.

\HI has also been detected in absorption against the CSS source with the WSRT \citep{Struve10c} at a low spatial resolution (\apx 10 kpc; Figure \ref{fig:vel_field}). CO (1-0) emission has also been detected from NGC 1167 \citep{Prandoni07, O'Sullivan15, Bolatto17}. The galaxy had originally been optically classified as a Seyfert 2 galaxy \citep{Ho97}. However, more recent observations show that the central optical AGN has a LINER (low ionization nuclear emission-line region) spectrum \citep[see ][]{Emonts_thesis}. NGC 1167 is also a part of the CALIFA survey and their integral field unit (IFU) studies further confirm the LINER nature of the nucleus \citep{Gomes16a}.

Here, we present high-resolution Very Large Array (VLA) and European VLBI Network (EVN) \HI absorption studies of the central CSS source B2 0258+35. We detect resolved, unusually broad \HI absorption that closely matches   the CO emission profile, pointing towards a circumnuclear structure that is being disturbed as a result of strong interaction with the expanding radio jets. 

In Section 2 we describe the VLA and EVN observations. We present the results in Section 3, discuss the implications of our results for studies seeking to quantify AGN feedback in Section 4, and provide a summary in Section 5.

Wherever required we have assumed a flat universe with H$_0$ = 67.3 km s$^{-1}$Mpc$^{-1}$, $\Omega_{\Lambda}$ = 0.685, and $\Omega_M$ = 0.315 \citep{Planck14}. At \textit{z}=0.0165, 1$''$ corresponds to 0.349 kpc.

\begin{table*}
%\centering
\begin{tabular}{ccccccccccc}
\hline
Telescope & $\nu_{\rm{obs}}$ & on-source & BW  & vel. resln & beam-size & PA & RMS$_{\rm{map}}$ & RMS$_{\rm{cube}}$ \\  & (GHz) & (hours) & (MHz) &(km s$^{-1}$) & $''\times''$ & ($^\circ$) & ($\mu$Jy beam\p{1}) & ($\mu$Jy beam\p{1}channel\p{1})\\ (1) & (2) & (3) & (4) & (5) & (6) & (7) & (8) & (9) \\ \hline

VLA  & 1.3965 & 5.85 &  6.25 & 21  & 1.13 $\times$ 0.98 & -75.11 & 340 & 840\\
EVN$^*$ & 1.4054 & 6.6  &  16   & 6.7 & 0.04923 $\times$ 0.04136 & 61.32 & 100 & 410 \\ \hline
\end{tabular}
\caption{Observation details: The columns are: (1) Telescope used; (2)  Central frequency of the band (GHz); (3)  On-source time in hours; (4) Bandwidth (MHz); (5) Velocity resolution (km/s); (6) Beam size; (7) Beam position angle ($^\circ$); (8)  RMS noise on the continuum map ($\mu$Jy beam\p{1}); and (9)  RMS noise on the spectral cube at a velocity resolution mentioned in (5) ($\mu$Jy beam\p{1}channel\p{1}).
\newline
$^*$Participating antennas: Effelsberg (Germany), Westerbork (the Netherlands), Jodrell-Bank (the United Kingdom), Medicina (Italy), Noto (Italy), Onsala (Sweden), Torun (Poland).}
\label{observations}
\end{table*}

\section{Observations and data reduction}

\subsection{Very Large Array observations}

The \HI observations with the VLA A-array were carried out in October 2008 (Proposal id: AS0955) for a total on-source time of 5.82 hours. A bandwidth of 6.25 MHz, centred at the redshifted \HI line frequency of 1396.5 MHz, subdivided into 64 channels was used as the set-up. This gave us a raw velocity resolution of 21 \kmpss. 3C147 was used for flux calibration and 0319+415 for phase and bandpass calibration. The details of the observations are listed in Table \ref{observations}. The observations were carried out when the VLA was being upgraded to the EVLA. The digital signal from the EVLA antennas had to be converted to analogue signals before being fed to the correlator, and this caused aliasing of power in the bottom 2 MHz of the band in all the EVLA-EVLA baselines. These baselines were flagged to avoid complications in solving for bandpasses and gains. Data were reduced in `classic' AIPS (Astronomical Image Processing Software). After the initial flagging of EVLA-EVLA baselines and bad data, antenna-dependent gains, and bandpass solutions were determined using the observations on the calibrators. The gain solutions were   further improved iteratively through the self-calibration procedure (imaging and phase-only self-calibration cycles until the continuum map showed no further improvement) after which a round of amplitude and phase self-calibration and imaging was done. Then the continuum model was subtracted from the calibrated visibilities and the residual UV data affected by radio frequency interference (RFI) were flagged. The final continuum model was subtracted from the calibrated multi-channel UV dataset. Residual continuum emission was removed by fitting a linear polynomial to each visibility spectrum. Finally, the continuum-subtracted data were shifted to the heliocentric frame. This dataset was then imaged to obtain the spectral cube.

The continuum map has an RMS noise of \apx 340 $\mu$Jy beam\p{1} and a restoring beam of 1.11$''$ $\times$ 1.02$''$ with a position angle of -75.11$^\circ$. This was made using ROBUST=-2 weighting, averaging all the line-free channels together. The peak flux density of the target is 1.08 $\pm$ 0.05~Jy~beam\p{1}. The integrated flux density is 1.97 $\pm$ 0.10 ~Jy. The uncertainty on the flux density scale at the observed frequency is assumed to be 5\pc \citep{Perley2017}. 
The continuum-subtracted spectral cube has the same restoring beam as the continuum map. The noise on the cube is \apx 840 $\mu$Jy beam\p{1} channel\p{1} for a channel width of 21 \kmpss, without any spectral smoothing.

\subsection{EVN observations}

The phase-referenced \HI observations with the European VLBI Network (EVN) were carried out in October 2012 (Project ID: ES070) for a total on-source time of 6.6 hours. We used a bandwidth of 16 MHz subdivided into 512 channels. Data reduction was carried out using `classic' AIPS. The details of the observations are listed in Table \ref{observations}.

We corrected for the instrumental delay, and then for delay and rate as a function of time after initial flagging of bad data, carried out bandpass calibration, and applied the solutions to the target. We then self-calibrated the target dataset to further improve the gain solutions. The self-calibration iterations are similar to those performed for the VLA data, but with the solution interval being shortened progressively from 30 minutes to 2 minutes for phase-only self-calibration. We applied these gain solutions to the UV data, subtracted the final continuum model from the calibrated UV data, and then flagged the residuals for RFI. We subtracted residual continuum emission by fitting a second-order polynomial to each visibility spectrum and imaged the residual UV data to obtain a spectral cube after shifting to the heliocentric frame. 

The final continuum map (obtained with natural weighting ROBUST = 5 and averaging all the channels together) has an RMS noise of 100 $\mu$Jy beam\p{1} and a beam of 49.23~mas~x~41.36~mas with a position angle of 61.32$^{\circ}$. The peak flux of the target is 95 mJy beam\p{1} and the integrated flux is 193 mJy. This was also estimated from the AIPS task JMFIT with a single component Gaussian fit.

The spectral cube was also obtained with natural weighting (ROBUST~=~5) and with the same restoring beam as the continuum map. The cube has an RMS noise of 0.4 mJy beam\p{1} channel\p{1} for a channel resolution of 6.7 \kmpss. We also tried other weighting schemes (uniform and robust) and found that the \HI absorption (see Section \ref{results}) was detectable only in the spectral cube obtained with natural weighting.

\begin{figure*}
%    \includegraphics[width=\columnwidth,height=7cm]{all_spectra}
    %\includegraphics[height=12cm]{vla_abs}
    \includegraphics[height=7.1cm, width=8.8cm]{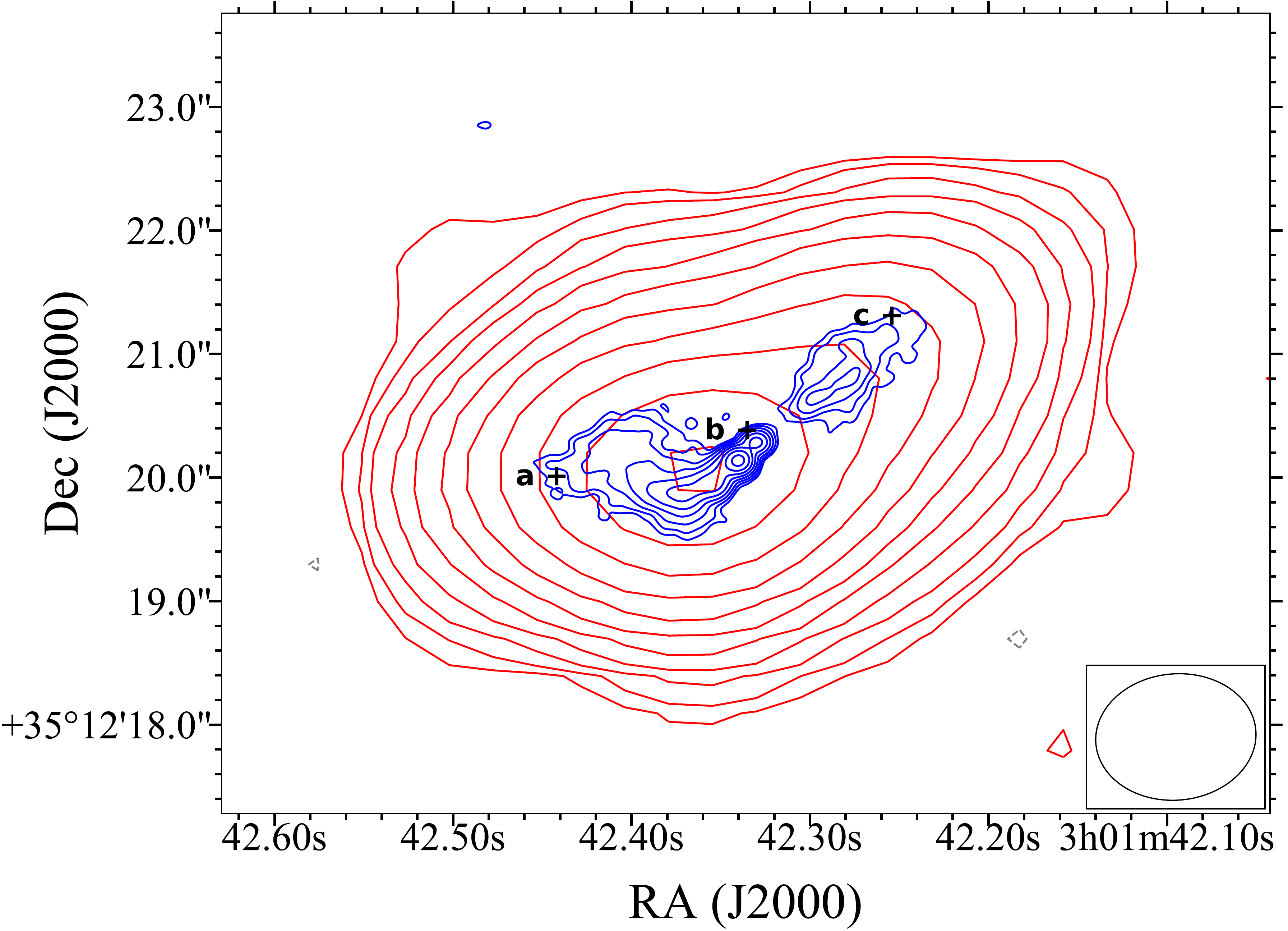}    
     \hspace{1em}
    \includegraphics[height=7.7cm, width=9.5cm]{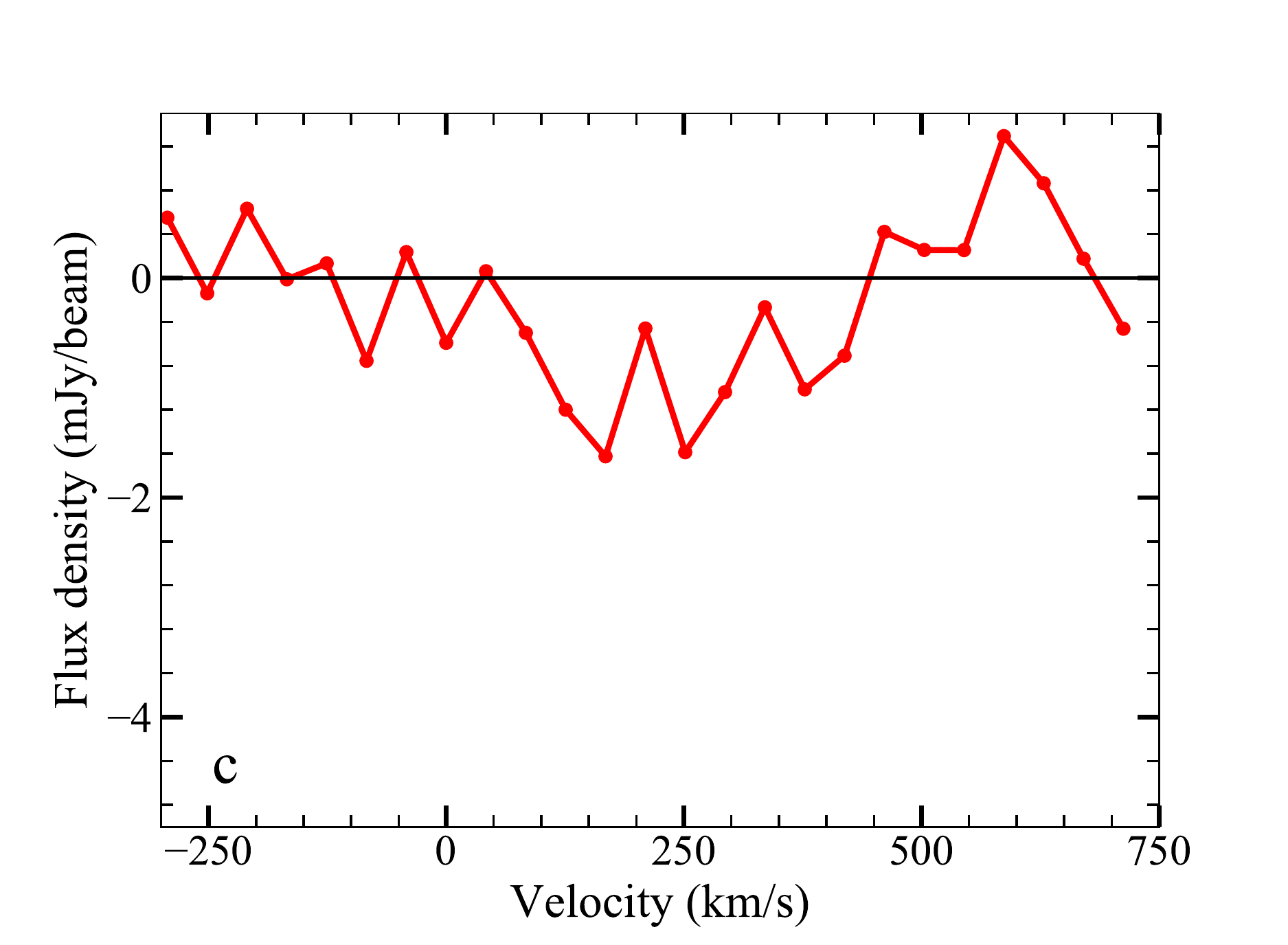}    
    \includegraphics[height=7.7cm, width=9.5cm]{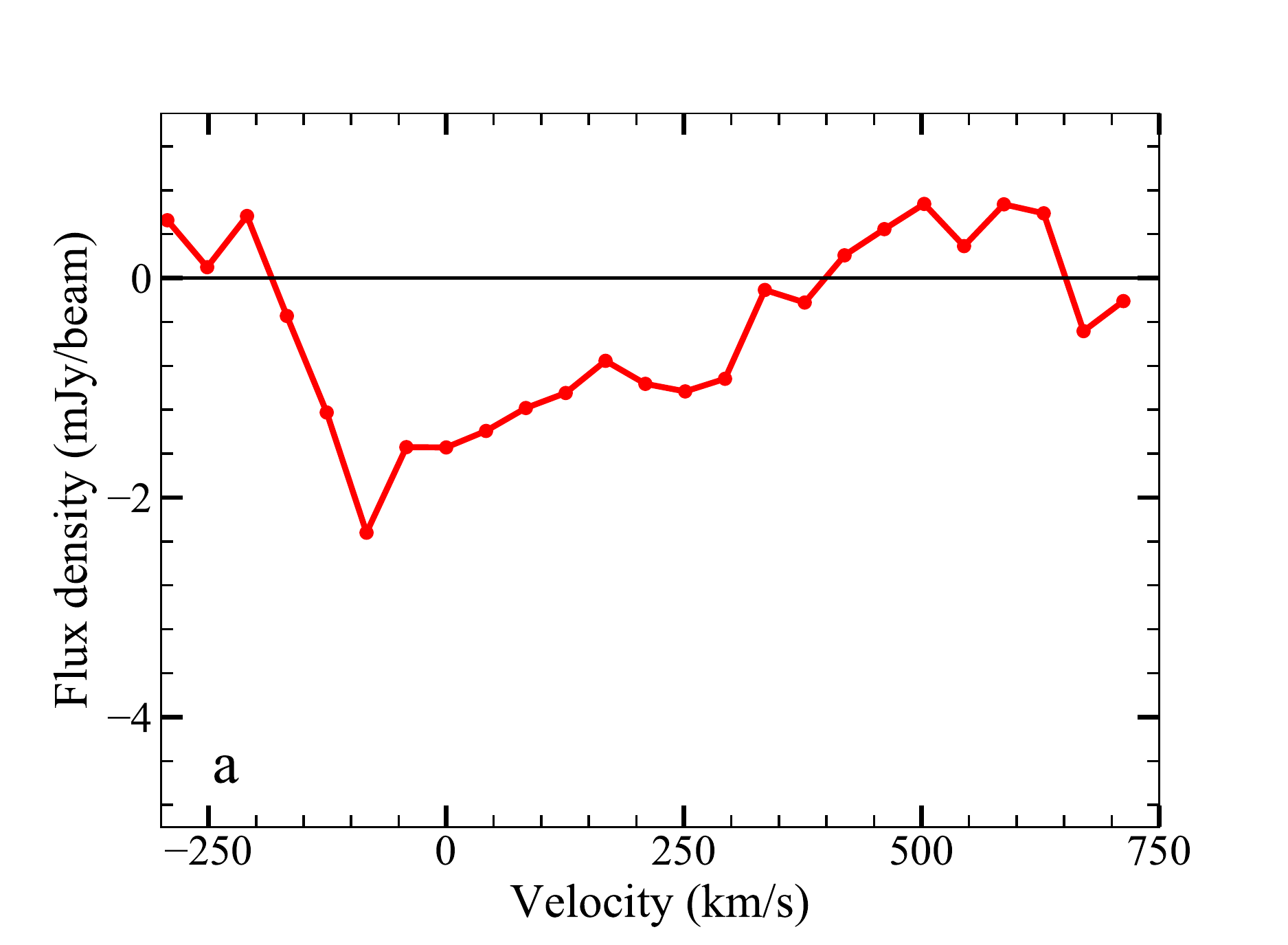}    
     \hspace{-0.9em}
    \includegraphics[height=7.7cm, width=9.5cm]{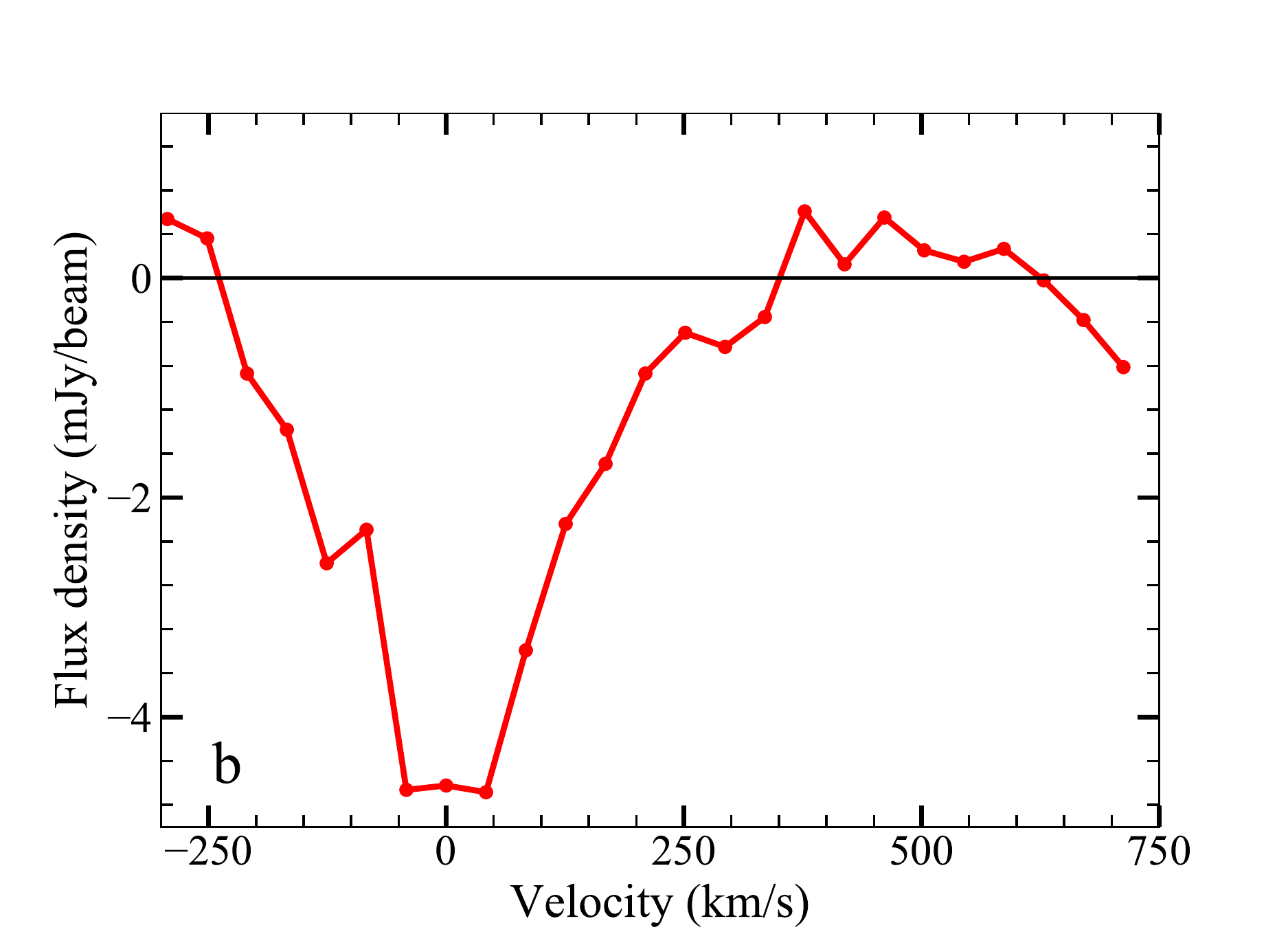}    
    \caption{\textbf{Top left:}  VLA continuum map of \target ( beam size: 1.13$''$ $\times$ 0.98$''$; red contours). The contour levels start from 1 mJy beam\p{1} (3$\sigma$) and increase by a factor of two. The 3$\sigma$ negative contours are shown in grey and the beam is shown in the bottom left corner of the image. The regions from which the \HI absorption spectra are extracted are labelled. The 22 GHz VLA map \citep[by][beam size: 0.12$''$ $\times$ 0.12$''$]{Giroletti05} is overlaid in blue contours to show the underlying continuum structure. These contours start from 325 $\mu$Jy beam\p{1} (5$\sigma$) and increase by a factor of two. \textbf{Bottom left, bottom right, and top right}: The absorption spectra extracted from the regions are  in black in the continuum map. The spectra have been Hanning-smoothed and resampled to a velocity resolution of \apx 42 \kmpss.}
    \label{fig:all_spectra}
\end{figure*}

\section{Results} \label{results}

\subsection{Continuum structure} \label{radio_continuum}

Figure~\ref{fig:all_spectra} shows the VLA continuum map of \target with the higher resolution 22 GHz map of \citet{Giroletti05} overlaid. In the VLA map the source is slightly resolved and is about 1 kpc in size. 
Higher resolution observations \citep{Sanghera95, Giroletti05} have shown that \target has plume-like lobes. In addition, there is a bright knot \apx 0.1$''$ south-east of the core. Further to the south-east the southern jet undergoes a sharp bend. There is no bright spot or bending in the northern jet. This asymmetry between the northern and the southern jets is reflected in our VLA map at a spatial resolution of \apx 0.35 kpc. We note that the integrated flux density from our observations is consistent with that from the WSRT observations at a resolution of \apx 10.1 kpc \citep[1.8 Jy;][]{Struve10c} within the flux measurement errors. Thus, there is no significant radio continuum outside the central \apx 1 kpc region.

\begin{figure}
    \includegraphics[width=\columnwidth]{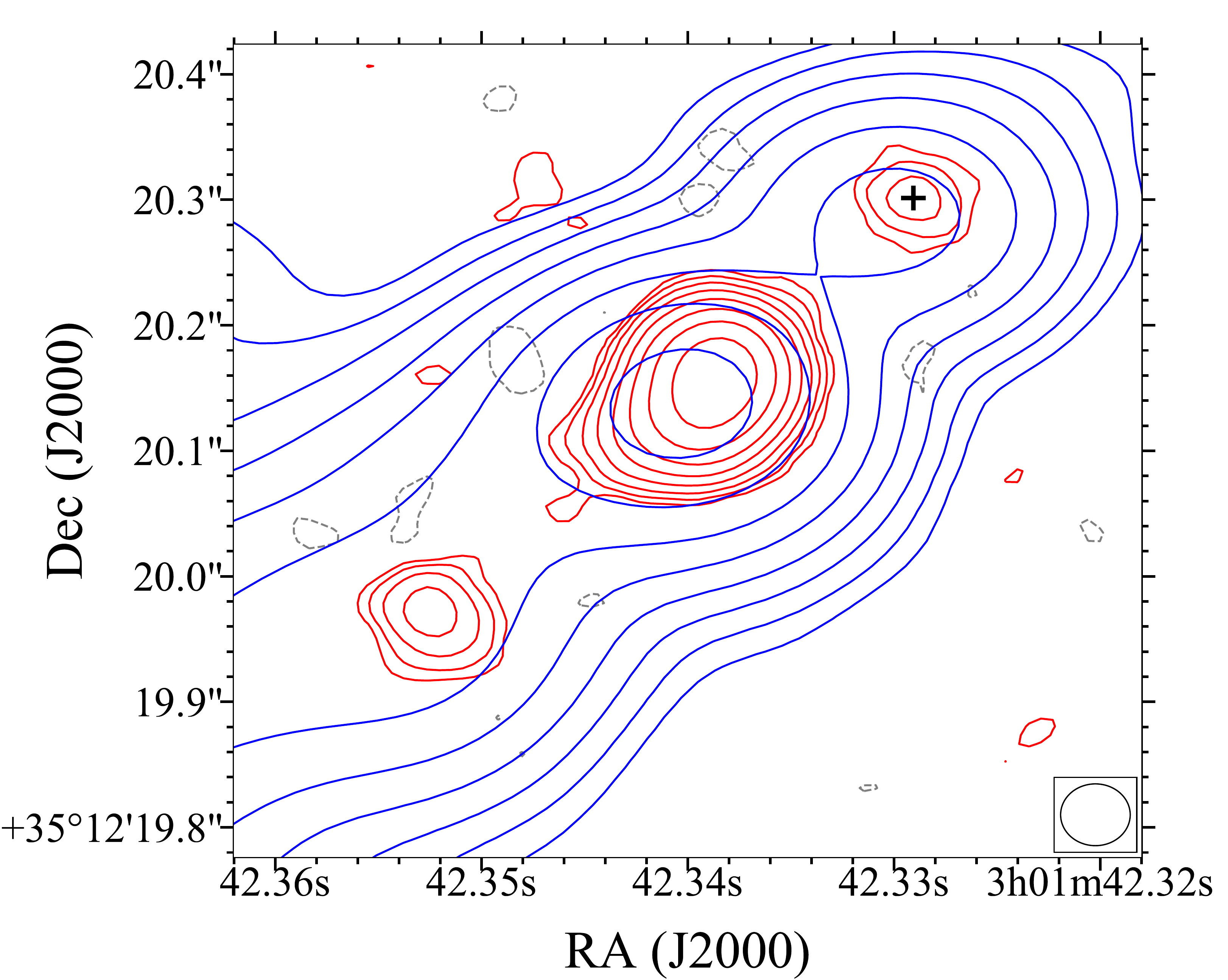}
    \caption{The VLBI map of \target (beam size: 0.04923$''$ $\times$ 0.04136$''$; red contours). The contours start at 400 $\mu$Jy beam\p{1} (4$\sigma$) and increase by a factor of two at each step. The 4$\sigma$ negative contours are shown in grey and the beam is shown in the bottom right corner of the map. The black cross marks the VLBI core. The 22 GHz VLA image from \citet{Giroletti05} is overlaid in blue contours for reference,  as in Figure \ref{fig:all_spectra}. Only 10\pc of the total VLA flux is recovered.}
    \label{fig:vlbi_22}
\end{figure}

With the EVN we recover only three components (see  Figure~\ref{fig:vlbi_22}): the faint unresolved core (with an integrated flux density of \apx 2.5 mJy), the bright extended knot (\apx 185 mJy)  coincident with the peak of the VLA map and  about 75 mas in size, and another knot \apx 0.2$''$ south-east of the bright knot (\apx 5 mJy), also resolved with an angular size of \apx 55 mas. We do not detect any jet-like structures connecting the core and this bright spot. The total integrated flux density is \apx 193 mJy, corresponding to about 10\pc of the flux density recovered with the VLA observations. Thus, the rest of the continuum flux is diffuse and distributed over length scales larger than the largest recoverable angular scale of our EVN array, \apx 200 mas (which corresponds to 80 pc). \target has also been observed with the VLBA at 1.6 GHz by \citet{Giroletti05}. They report the detection of the bright knot and the core, but not the third component to the south of the bright knot. They measure a total flux density of 243 mJy over the entire radio source, consistent with our measurement within the flux measurement errors. However, the peak flux density we measure is an order of magnitude higher than their measurement. These differences could be due to the much higher spatial resolution (a beam size of 10.4 mas $\times$ 4.3 mas) of their observations.

\subsection{H \small{I} \normalsize{absorption}} \label{hi_absorption}

We detect slightly resolved \HI absorption against the radio continuum in our VLA data. The absorption profiles against different parts of the continuum are shown in Figure~\ref{fig:all_spectra}. Since the continuum is only a few beams across, we  extracted the spectra from regions sufficiently far apart so that each spectrum extracted is from a different beam element: against the northern jet, the region including the core and the bright knot, and the southern jet. We see strong absorption against the continuum in the beam elements covering the bright knot, and the southern jet. We also detect weak absorption against the northern jet. The full width at zero intensity (FWZI) of these \HI absorption profiles is \apx 545 \kmps against the southern jet and the hotspot, while it is \apx 500 \kmps against the northern jet. The column density of the gas detected at the locations shown in Figure \ref{fig:all_spectra} are consistent with each other within the measurement errors; they range from (3 $\pm$ 0.6) $\times$ 10\pp{20} cm\p{2} against the northern jet to (2.2 $\pm$ 0.1) $\times$ 10\pp{20} cm\p{2} against the hotspot. This was estimated by assuming a spin temperature of 100 K and a covering factor of unity.

\begin{figure}
    \includegraphics[width=\columnwidth]{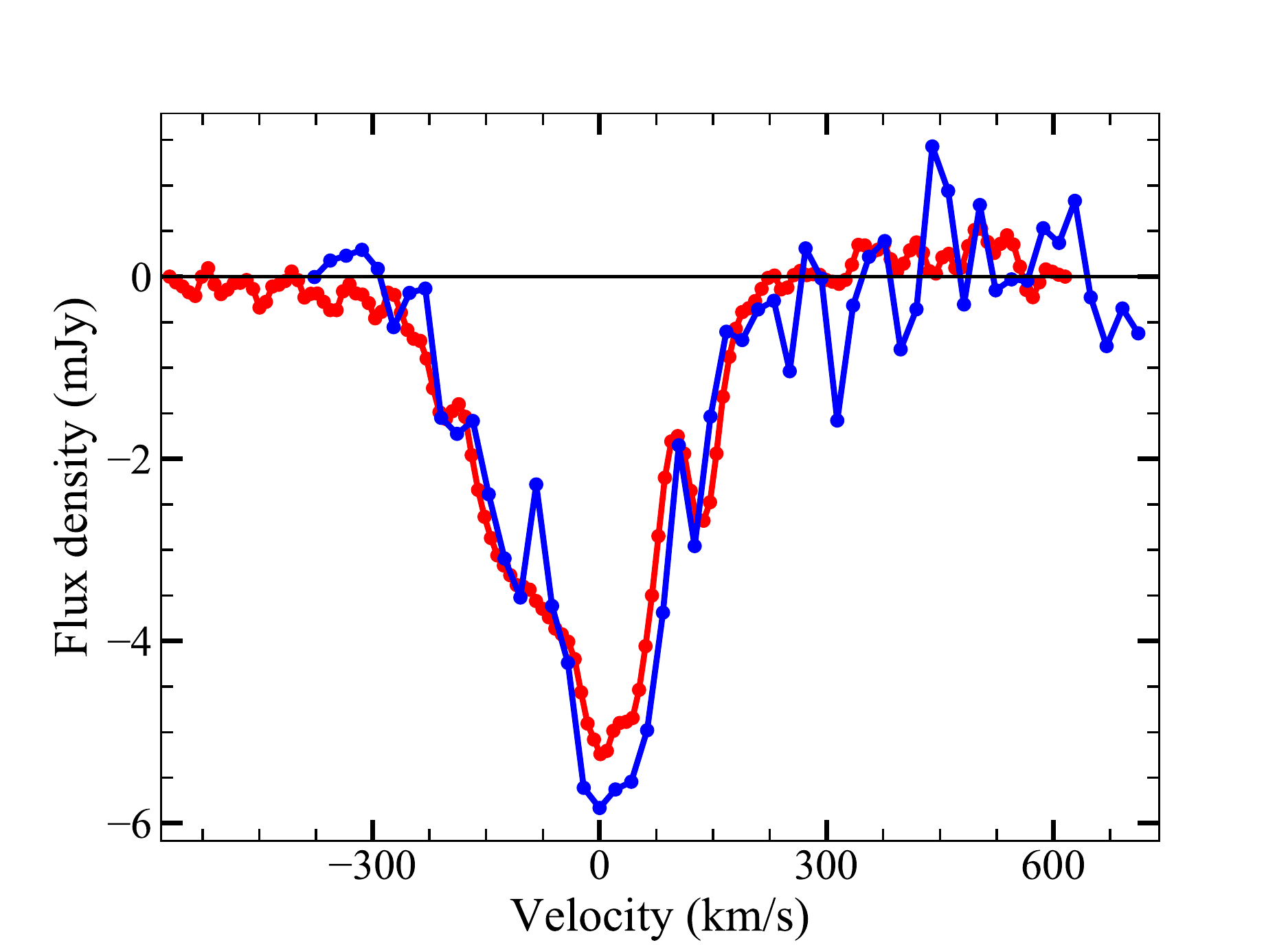}
    \caption{Integrated VLA (blue) and WSRT (red) \HI absorption profiles. The profiles are in excellent agreement with each other, implying there is no diffuse component lost in continuum and in absorption.}
    \label{fig:vla_wsrt}
\end{figure}

In addition, Figure~\ref{fig:vla_wsrt} shows a comparison between the integrated VLA and the WSRT \HI absorption profiles \citep{Struve10c}. Given the smaller bandwidth of the VLA observations and the higher noise level, the two profiles are in good agreement with each other. Thus, the higher spatial resolution of the VLA observations does not result in non-detection of a significant fraction of \hi.

\begin{figure}
    \includegraphics[width=\columnwidth]{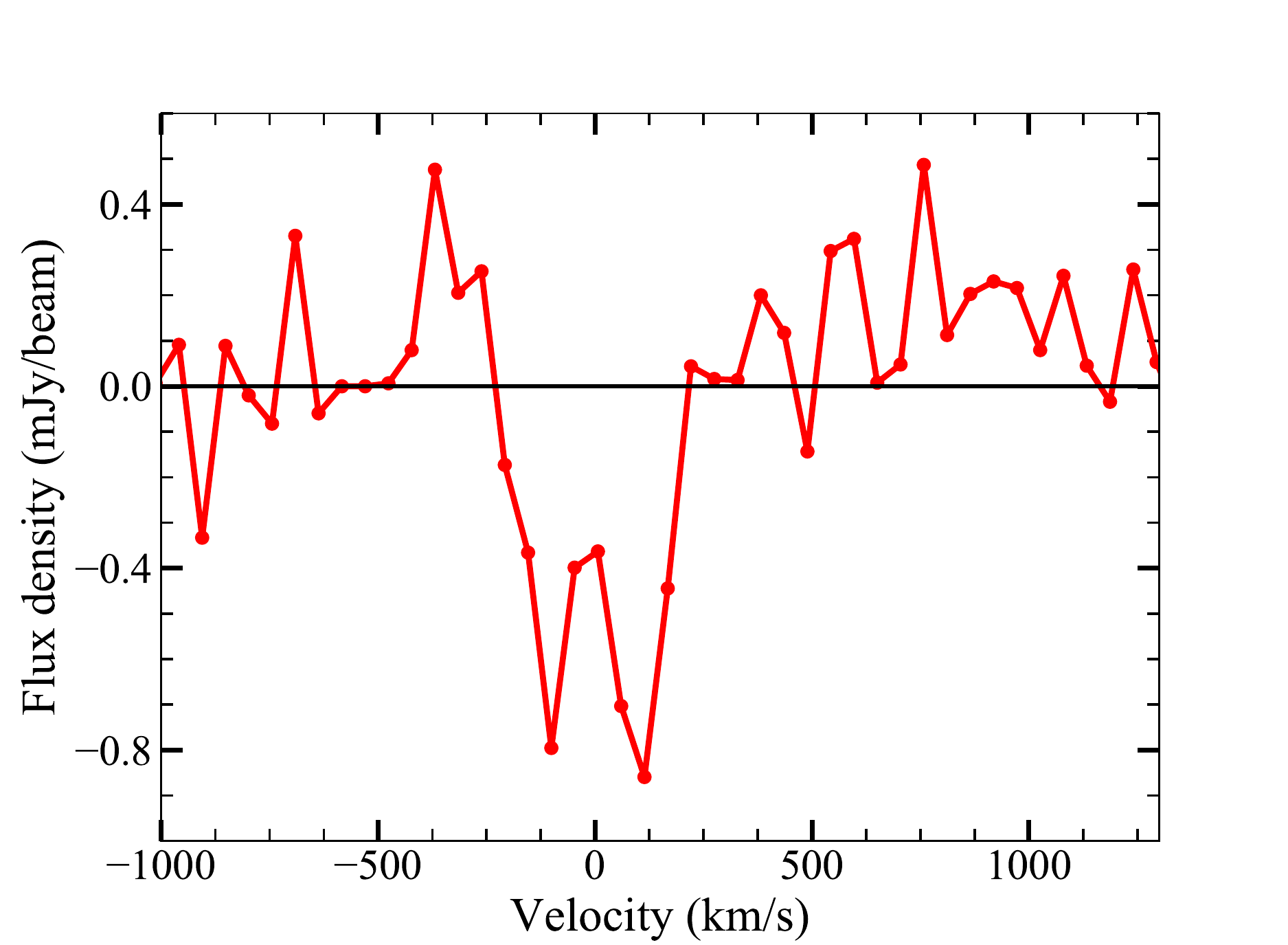}
    \caption{Integrated \HI absorption spectrum against the knot (Figure ~\ref{fig:vlbi_22}) recovered with the EVN. The spectrum has been Hanning-smoothed to a velocity resolution of 54 \kmpss. The RMS noise on the Hanning-smoothed spectrum is 30 $\mu$Jy and the FWZI of the absorption feature is \apx 322 \kmpss.}
    \label{fig:vlbi_hi}
\end{figure}

With the EVN observations, we do not detect any absorption at the location of the peak of the knot or the southern blob or against the core. However, we do find an absorption feature when integrated over the region of the knot (see  Figure ~\ref{fig:vlbi_hi}). The absorption profile has an FWZI of \apx 322 \kmps and a peak absorption of -0.85 mJy. Since a considerable amount of the continuum flux density is resolved out at the VLBI scale, the peak of the recovered absorption is \apx 5 times lower than that of the VLA absorption spectrum. However, the \HI column density derived is (1.70 $\pm$ 0.21) $\times$ 10\pp{20} cm\p{2}, which is consistent with the column density measured against the knot from the VLA spectrum. This detection thus implies a uniform coverage of the radio continuum by the absorbing gas and hence a covering factor of unity.

\section{Discussion} \label{discussion}

With the information collected, we first wanted to investigate whether the \HI seen in absorption is a part of the large-scale disc and is distributed over a few tens of kpc or is co-spatial with the jet and if so, whether there is any signature of jet--ISM interaction. We did this by comparing the absorption profile with a model absorption profile arising from the large-scale  disc and then combining this with the results from molecular gas emission.

\subsection{Origin of H \small{I} \normalsize{absorption}}\label{modelling}

The shape of the absorption profile, in general, depends  on the distribution of gas and on the morphology of the background radio source. As  Figure \ref{fig:vel_field} shows, the large-scale \HI disc has a position angle of 75\pp{\circ} and the approaching side of the disc is towards the knot and the southern jet of \target. The radio continuum in our case is asymmetric with the southern knot;  the southern jet is considerably brighter than the northern jet (see Section \ref{radio_continuum}) and has a position angle of 132\pp{\circ}. If the absorption solely arose from \HI belonging to this large disc, the southern lobe would give rise to much stronger absorption compared to the northern lobe. Thus, the integrated absorption profile would be asymmetric with stronger absorption towards the blueshifted velocities. Furthermore, the absorption profile towards the southern jet (for example, at position `a' shown in Figure \ref{fig:all_spectra}) should have a blueshifted wing instead of a redshifted one, unlike what is observed. To produce a symmetric absorption profile, the absorbing gas and hence also the radio jets should be along the minor axis of the rotating disc, which is not the case here (see Figure \ref{fig:vel_field}).

This can be illustrated by modelling the \HI absorption arising from an \HI disc around the radio source, in this case the galactic \HI disc. The parameters involved in this modelling are the rotation curve of the disc, inclination and position angles, and the velocity dispersion of the gas.

\begin{figure*}
    \includegraphics[width=\columnwidth]{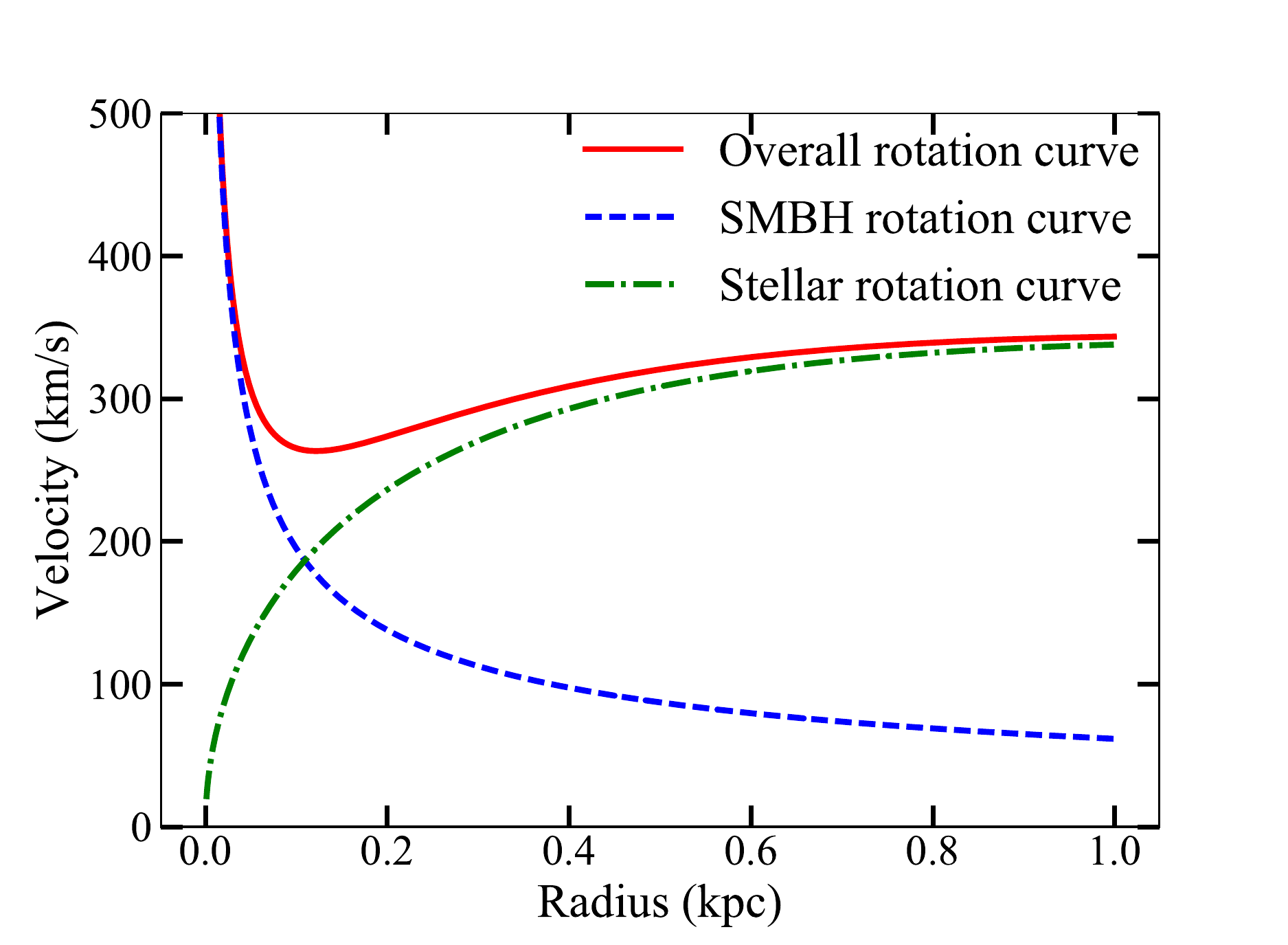}
    \includegraphics[width=\columnwidth]{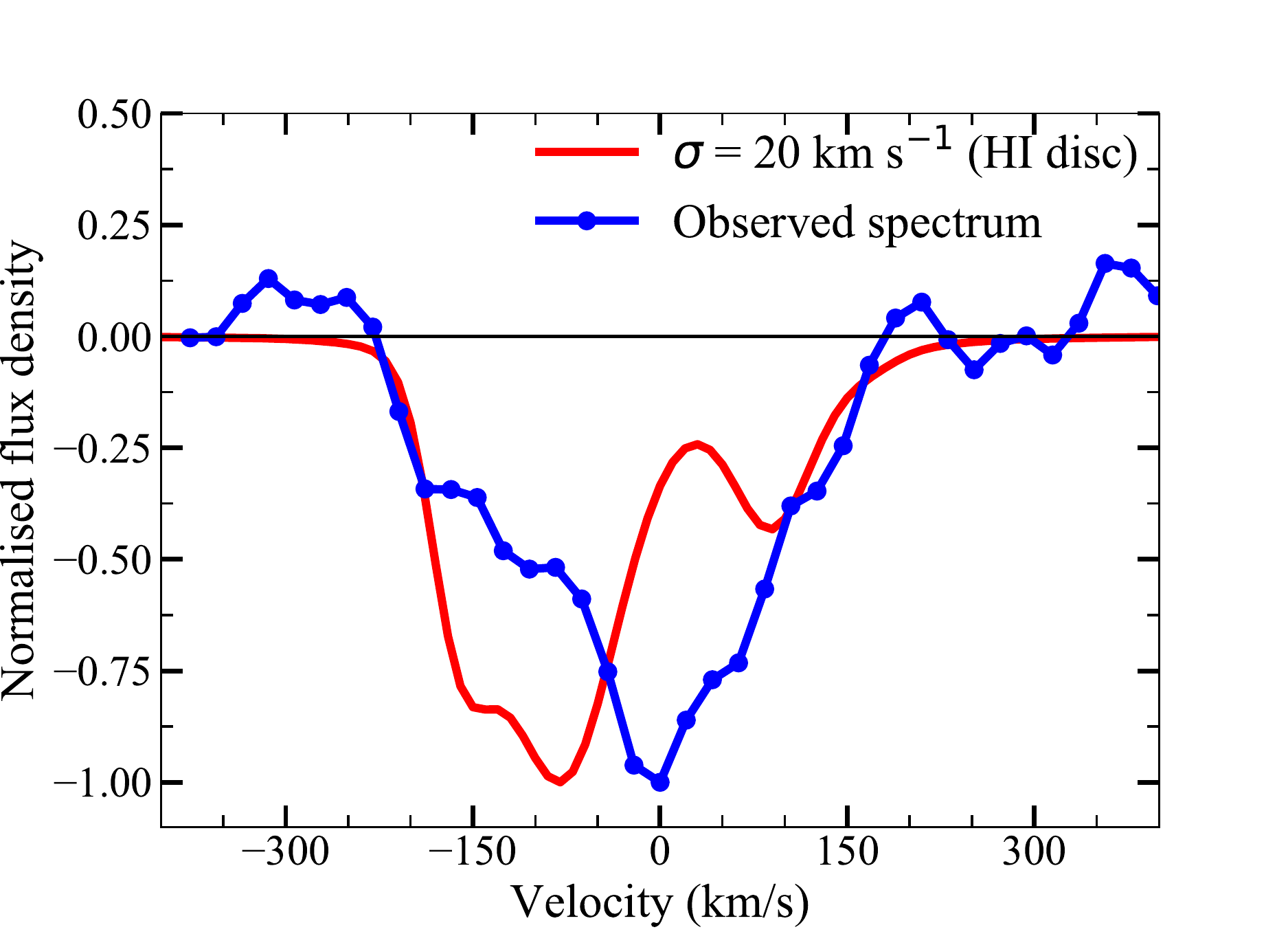}
    \caption{\textbf{Left panel:} Rotation curve for NGC 1167 in the inner 1 kpc. In the inner regions the major contributions to the rotation curve come from the stellar component and the supermassive black hole. \textbf{Right panel:} Observed absorption profile (blue) and the model absorption profile (in red) for the large-scale \HI disc.}
    \label{fig:model_1}
\end{figure*}

In the nuclear regions of the galaxy, the major contribution to the rotation curve comes from the stellar component of the galaxy and the SMBH. The SMBH mass of NGC 1167 is 4.4~$\times$~10\pp{8}~M$_{\odot}$ \citep{Kormendy13}. For a stellar velocity dispersion of 219.6 \kmps \citep{Ho09} the black hole sphere of influence extends up to \apx 40 pc. To estimate the contribution to rotation by the stellar component, we  used the \citet{Hernquist90} model with an effective radius of 6.7$''$ and a total stellar mass of 1.3 $\times$ 10$^{13}$ M$_{\odot}$ as parameters \citep{Noordermeer07}.  The resulting rotation curve in the inner 1 kpc region of NGC 1167 is shown in Figure \ref{fig:model_1}, left panel. We note that the rotation curve is essentially flat at 300 \kmps even in the very central region of the galaxy. 

The inclination (38\pp{\circ}) and the position angles (75\pp{\circ}) of the disc have been well constrained by \citet{Struve10c}. In addition, we  assume a velocity dispersion ($\sigma$) of 20 \kmpss, a typical value observed in the central regions of early-type galaxies. However, \citet{Struve10c} have found that a model with a velocity dispersion of 9 \kmps explains the \hi\ rotation curve of NGC 1167 very well. Hence, our assumption of 20 \kmps for the velocity dispersion is quite conservative. The resulting absorption profile is shown in Figure. \ref{fig:model_1}, right panel. The profile is asymmetric (as explained earlier), unlike the observed profile and does not match the observed location of the peak absorption. This supports our hypothesis that the absorbing \hi\ has kinematics that is quite different from the large \HI disc.

An extreme possibility that can give rise to an absorption profile centred at the systemic velocity is that the gas does not have any rotational motion, but is entirely chaotic. Another possibility that can lead to a symmetric \hi\ absorption profile is that the disc has undergone a warp in the central few kiloparsecs of the galaxy with an orientation such that the kinematical minor axis in the inner region coincides with the jet axis. In this case all absorption should appear close to the systemic velocity and the large width of the absorption would also imply a large dispersion of \HI.

However, the resolved \HI absorption spectra shown in Figure \ref{fig:all_spectra} do show a sign of rotation with the absorption centroid shifting redwards as we move towards the northern jet. Thus, we argue that the absorption arises from a kiloparsec-scale gas disc with a velocity dispersion higher than that of the large \hi\ disc (further supported by the nature of CO emission; see the following subsection) whose orientation is different from that of the large \HI disc.

\subsection{Comparison with molecular gas} \label{CO}

In order to further strengthen this hypothesis about the morphology and kinematics of the absorbing \hi, we make use of the available information on molecular gas. Observed in emission, the distribution of gas traced by the molecular component is independent of the size and asymmetry of the background continuum. Thus, when combined with the absorption studies, it can provide further insights into the kinematics and morphology of the gas under investigation.

CO(1-0) emission has been detected from NGC 1167  by many studies: \citet{Prandoni07}, \citet{O'Sullivan15} and \citet{Bolatto17}. The first two studies have made use of IRAM single-dish observations, while the last one is an interferometric study with the Combined Array for Research in Millimeter-wave Astronomy (CARMA). The single-dish observations detect CO(1-0) emission from the central \apx 8 kpc (23$''$; the resolution of the telescope) region of the galaxy. The interferometric observation of \citet{Bolatto17} has a larger field of view, but is of poorer sensitivity. They report the detection of very faint CO emission from the region outside the IRAM field of view, but are not sensitive to the emission seen by IRAM in the central part of the galaxy. Furthermore, the CO detected by \citet{Bolatto17} is coincident with the low-level star formation sites in NGC 1167 reported by \citet{Gomes16}, and thus  we only consider the deeper IRAM observations of CO in the central region for our discussion.

\begin{figure*}
    \includegraphics[width=20cm]{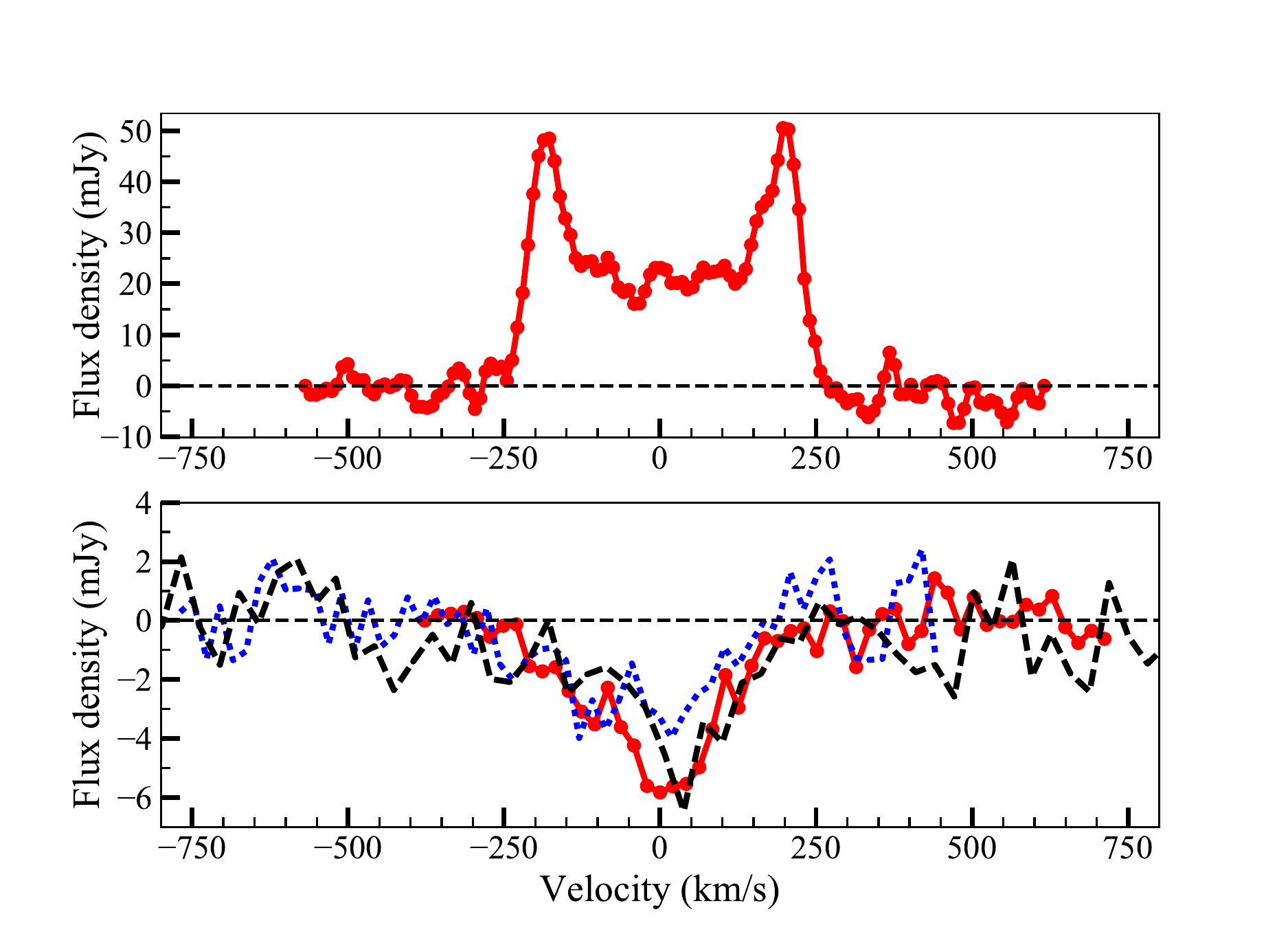}
    \caption{Comparison between the \HI emission profile \citep[top panel, red:][]{Struve10c}; CO emission profiles, inverted and in arbitrary units for  ease of comparison (bottom panel, blue dotted: \citealt{Prandoni07}; black dashed: \citealt{O'Sullivan15}); and the WSRT \HI absorption profile (bottom panel, red). The CO emission profile does not match  the \HI emission profile, implying that the CO emission does not arise from the molecular counterpart of the large \HI disc. The \HI absorption and the CO emission profiles match, implying that CO and absorbing \HI are spatially coincident.}
    \label{fig:ems_abs}
\end{figure*}

The IRAM emission spectra presented by \citet{Prandoni07} and \citet{O'Sullivan15} agree with each other in their line widths (FWHM of 261 $\pm$ 36 \kmps and 233 $\pm$ 47 \kmps respectively) and  in the integrated emission: 0.93 $\pm$ 0.14 K \kmps and 1.18 $\pm$ 0.1 K \kmps respectively (shown in Figure \ref{fig:ems_abs}, bottom panel).

The integrated CO(1-0) emission profile is shown in the bottom panel of Figure~\ref{fig:ems_abs}. The profile is inverted and plotted in arbitrary units for ease of comparison with the \HI absorption profile. As mentioned earlier, the rotation curve for the galaxy is approximately flat at 300 \kmps even within the inner kiloparsec region (Figure \ref{fig:model_1}, left panel). Thus, we would expect the CO emission profile to be double-horned if it formed a regularly rotating disc. This is  not the case, which suggests that the kinematics of CO is not dominated by rotation. This supports our earlier hypothesis of the presence of a kiloparsec-scale turbulent gas disc. Assuming the gas to be entirely chaotic, we find an upper limit to the velocity dispersion of CO to be \apx 90 \kmpss, estimated via a single-component Gaussian fit to the emission profile of \citet{O'Sullivan15}, which has a higher signal-to-noise ratio compared to that of \citet{Prandoni07}. 

Next, we find (as can be seen in Figure \ref{fig:ems_abs}) that the line widths of the CO emission and the \HI absorption profiles match, as do  the natures of the two profiles. This suggests that these two components of gas (molecular and atomic) are spatially coincident, and hence CO is distributed within the central few kiloparsecs of the host galaxy. Thus, this spatial coincidence  suggests that the \HI seen in absorption is also dominated by turbulence rather than bulk rotation. In addition, the covering factor of the gas is unity since we detect \HI absorption against the entire radio source. We thus posit that the gas seen in \HI absorption and CO emission forms a kiloparsec-scale disc into which the radio jets are expanding and injecting turbulence, causing the observed line profiles. 

The line ratio CO(2-1)/CO(1-0) can also be used to investigate the presence of turbulence. This line ratio, as reported by \citet{O'Sullivan15}, is unity, which is twice the expected value (of 0.5) for quiescent molecular gas in normal ISM \eg{Penaloza17} and similar to the value observed in other cases of strong jet--ISM interaction \citep[for example IC 5063;][]{Oosterloo17}. This further supports the presence of strong turbulence in the medium. 

The presence of turbulence has been observed in other objects for example, in 3C326N \citep{Guillard12} and in NGC1266 \citep{Alatalo15a}. Following a similar approach to that used for these sources, we estimated the turbulent component of energy associated with the gas. The  turbulent kinetic energy is given by E$_{\rm{turb}}$ = 3/2 M$_{\rm{mol}} \sigma^2$ \citep[e.g.][]{Guillard12}. The molecular gas is likely to be more massive than the atomic gas in the nuclear regions and hence will contain most of the turbulent energy. We estimate this energy associated with the molecular phase by assuming a velocity dispersion ($\sigma$) of 90 \kmpss. We would like to note here that the value of velocity dispersion was derived assuming the gas to be entirely chaotic with no systematic motion. The turbulent energy estimated in this way will thus be an upper limit. To estimate the molecular gas mass, we assume a range of values for the CO to H$_2$ conversion factor (X$_{\rm{CO}}$): 0.2 $\times 10^{20}$ (K \kmpss)$^{-1}$, a conservative value representing optically thin turbulent gas;  0.4 $\times 10^{20}$ (K \kmpss)$^{-1}$ representing dense gas in ultra-luminous infrared galaxies; and 2 $\times 10^{20}$ (K \kmpss)$^{-1}$, the Galactic value \citep{Bolatto13}. The corresponding molecular gas masses are 8.8 $\times$ 10$^{6}$ M$_{\odot}$, 1.8 $\times$ 10$^{7}$ M$_{\odot}$, and 8.8 $\times$ 10$^{7}$ M$_{\odot}$. With these assumptions, we derive a turbulent kinetic energy (in the molecular phase) of 2.1 $\times$ 10$^{54}$ erg, 4.3 $\times$ 10$^{54}$ erg, and 2.1 $\times$ 10$^{55}$ erg, respectively, corresponding to the X$_{\rm CO}$ factors mentioned above.

\begin{figure}
    \includegraphics[width=\columnwidth]{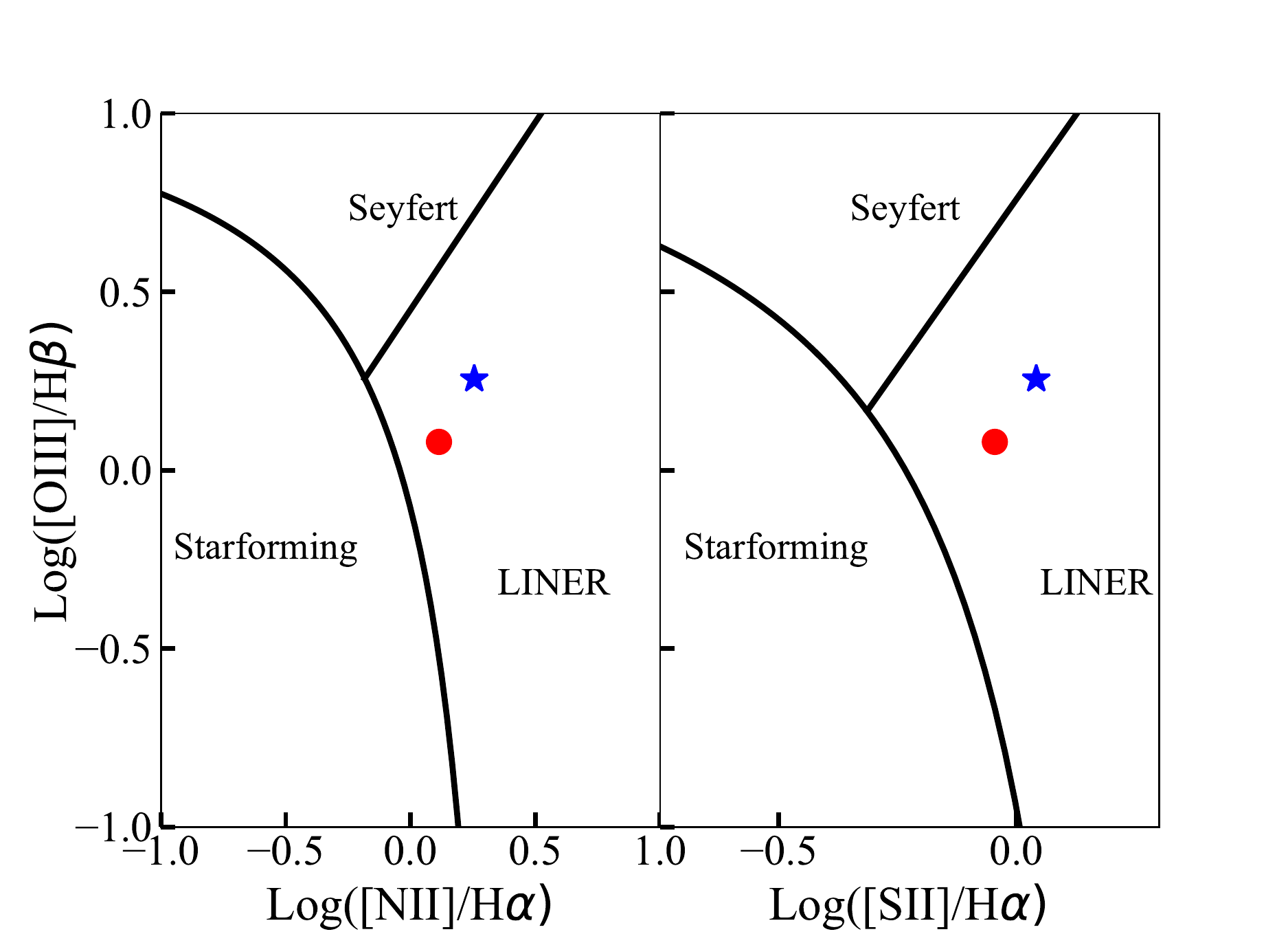}
    \caption{BPT diagrams showing the line ratios \citep{Emonts_thesis} of the off-nuclear regions of \target. The blue stars represent the line ratios from the spectrum extracted from the south-western region, offset by \apx 1.4 kpc from the nucleus, with PA=64\pp{\circ}. The red dots represent the line ratios from the spectrum extracted at a similar distance off the nucleus in the north-eastern region. The points clearly lie in the LINER portion of the diagram, suggesting the presence of shocks.}
    \label{fig:bpt}
\end{figure}

\subsection{Can jets cause this turbulence?}

In \targets, we find that both the \HI absorption and the CO emission profiles suggest that the gas is turbulent. 
We suggest that radio jets expanding through the gas are the main cause of this turbulence (see Sections \ref{modelling} and \ref{CO}). The effect of radiation or wind is unlikely to be significant since the optical AGN in \target is characterised by a LINER spectrum \citep{Gomes16a} with only weak emission lines. 

For  a scenario where  jets are expanding in the presence of a gas disc, the numerical simulations \eg{Mukherjee18b} predict that when the jets are perpendicular to the plane of the disc, they break out of the ISM quickly ($\lesssim$ 100 kyr). They also give rise to energy bubbles, which expand laterally into the gas disc generating radial gas outflows at velocities as high as \apx 500 \kmpss. The jets that are partially inclined to the gas disc or in the plane of the disc evolve much more slowly, and in the process also get deflected by the gas clouds. They too launch expanding bubbles in the gas disc, which evolve slowly. These simulations further predict that the mean velocity dispersion of the gas increases by several times its initial value, up to a few hundred kilometres per second, both at the point of direct interaction with the jet head and further out, as the energy bubble thus created spreads over the entire disc (e.g.   \citealt{Mukherjee18a},  section 3 (v)).

The jet power of \target is \apx 10\pp{44}~erg~s\p{1}, based on the scaling relation between cavity power (jet power) and radio luminosity \citep{Cavagnolo10}. Over a time span of \apx 0.4 Myr \citep[the lower limit on the age of the AGN;][]{Brienza18}, the jets are capable of depositing a total of \apx 10\pp{57} erg. Thus, the radio jets appear to be capable enough to provide the energy required to support the maximum level of turbulence even if only 1\pc of the total energy in the jet is converted to turbulent energy. However, we  note that the jet power estimation is quite uncertain owing to the large scatter of about an order of magnitude in the scaling relation and several uncertainties in estimating the energy losses. For example, in the case of another low-luminosity radio AGN IC~5063 \citep{Morganti15}, the jet power estimated using this prescription was shown to be an order of magnitude lower than that required to create the observed velocity dispersion in gas \citep{Mukherjee18b}. If the jet power  turns out to be higher, the efficiency required is even lower.

\subsection{Impact of the radio continuum on the host galaxy}\label{impact}

In order to understand the effect of \targets, which is \apx 1 kpc in size, on its host galaxy, we need to explore whether the impact of the radio jets is limited only to the central kpc region or extends to larger radii. Various multiwavelength studies have shown the effects of radio jets can indeed span distances much larger than the size of the radio jets.

\citet{Mahony16} found, using IFU spectroscopy of 3C 293, ionised gas outflows with linewidths $>$500 \kmps at 12 kpc from the nucleus along the radio axis and outflows with linewidths as high as 300 \kmps at distances as far as \apx 3.5 kpc perpendicular to the radio axis. Similarly, in 4C 31.04 \citep{zovaro19}, it was found that the jet-induced shocks extend to regions four times the size of the observed radio lobes. \citet{Rodriguez-Adrila17} report the presence of two expanding ionised gas shells that are spatially coincident with the radio jets. \citet{Fabbiano17, Fabbiano18b} suggest, via deep Chandra imaging and spectroscopy, that the radio jets of ESO 428-G014 give rise to \apx 2-3 kpc extended X-ray emission along the major axis of the radio source. In NGC 6764, the interaction between the parsec-scale radio jets and the ISM is thought to be one of the causes of subkiloparsec-scale radio bubbles \citep{Kharb10}. \citet{Maksym17} and \citet{Wang10c} also find evidence, via X-ray studies, of the jet--ISM interaction shaping the inner kpc region of the host galaxies of NGC 3396 and NGC 4151, respectively. 

Predictions from the numerical simulations of jets expanding at an inclination to a clumpy gas disc \eg{Sutherland07, Cielo18, Mukherjee18a, Mukherjee18b} agree with these findings. They predict that the jet flow will be channelled through the disc in different directions, which can ionise the gas even in the regions far from the radio jets. 

We can investigate the presence of such an effect in \target using the long-slit spectroscopy presented by \citet{Emonts_thesis}. In this study, the slit is aligned with the major axis of the host galaxy (PA = 64$^\circ$), far removed from the radio axis (PA=132\pp{\circ}). The spectra have been extracted at the location of the nucleus and at the off-nuclear regions 4$''$ south-east and north-west of the nucleus. Thus the off-nuclear spectra are extracted at a distance of \apx 1.3 kpc  from the nucleus, perpendicular to the radio axis. The [O III]/H$\beta$, [NII]/H$\alpha$, and [S II]/H$\alpha$ line ratios are plotted on diagnostic diagrams introduced by \citet{Baldwin81}, here referred to as BPT diagrams in Figure \ref{fig:bpt}. We see that the line ratios clearly fall in the low-ionisation nuclear emission-line region \citep[LINER;][]{Heckman80} part of the diagram. This is further confirmed by the IFU study of \citet{Gomes15b}, who  present a very detailed BPT diagram up to a diameter of \apx 30 kpc of NGC~1167. Except for the low surface brightness star-forming spiral arms, the remaining regions clearly fall in the LINER portion of the diagram.

LINER spectra can be explained by various mechanisms such as photoionisation either by nonstellar UV continuum or shocks, or evolved stars. From the IFU analysis, \citet{Gomes16a} identify two regions in the disc of NGC~1167 (see their Figure 1): the inner region ($\sim 4''$; 1.4 kpc) is characterised by high equivalent width H$\alpha$ emission, suggesting that evolved stars alone cannot be the source of ionisation, and the outer region with a lower equivalent width where the LINER spectrum can be explained by pAGB stars. \\
Thus, wide-spread shocks are the most plausible cause of the LINER spectra in the inner region. A major merger cannot be the reason for these shocks because such a possibility has already been ruled out by \citet{Struve10c}. We thus suggest that they are a result of the interaction between the radio jets and the ISM as described earlier.

It is more challenging  to ascertain whether this interaction has an effect on the rest of the disc, outside the inner few kpc. Interestingly, it is worth noting here that in a study of group dominant galaxies, \citet{O'Sullivan15} have found that NGC 1167 lies in the red sequence of quenched galaxies. However, with the present data, it is not possible to ascertain whether this could have, at least partly, resulted from the AGN activity. High-resolution mapping of molecular and ionised gas may shed further light in this direction.

\begin{figure*}
    \includegraphics[width=12cm, height=10cm]{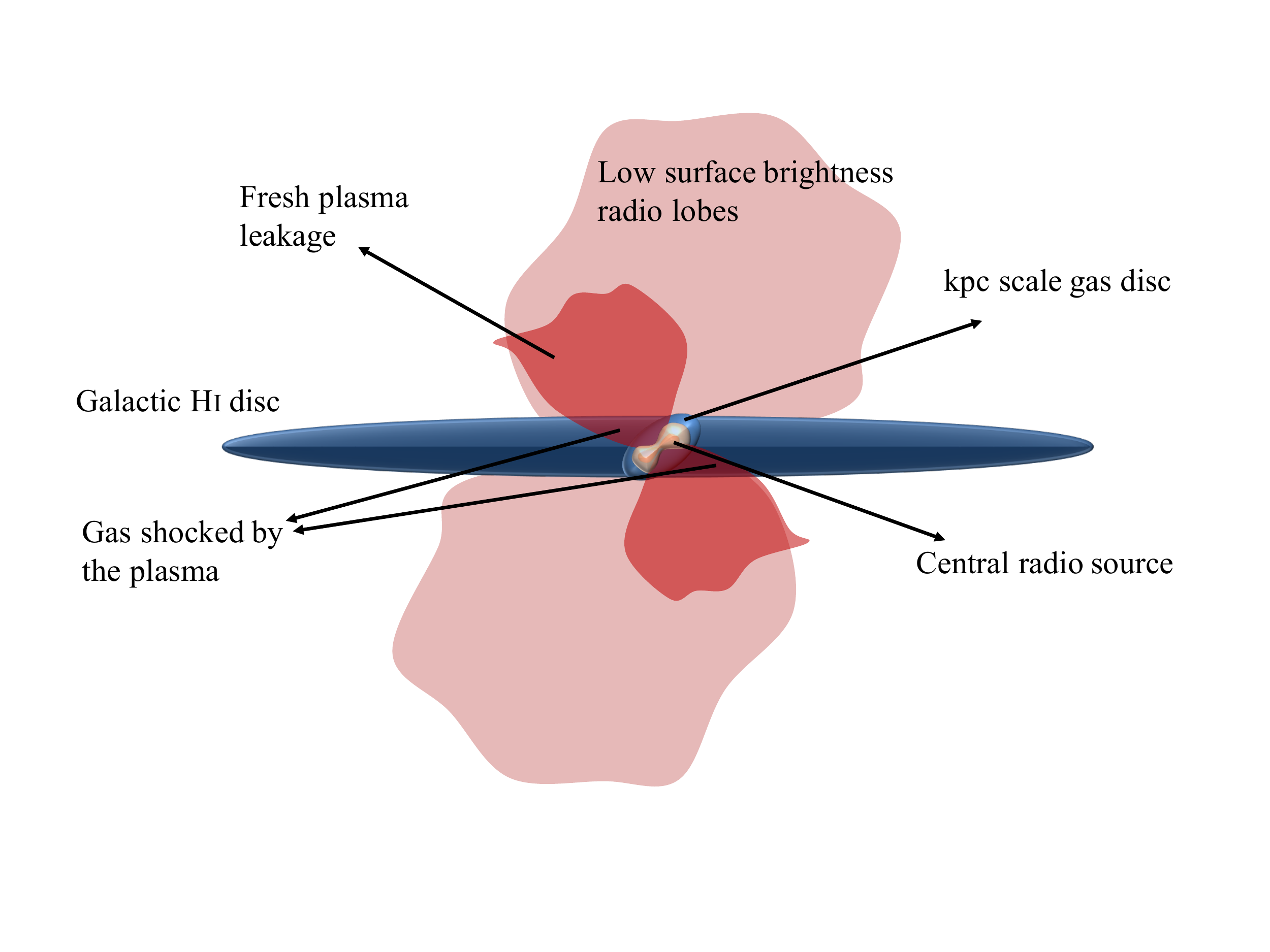}
    \includegraphics[width=6.7cm]{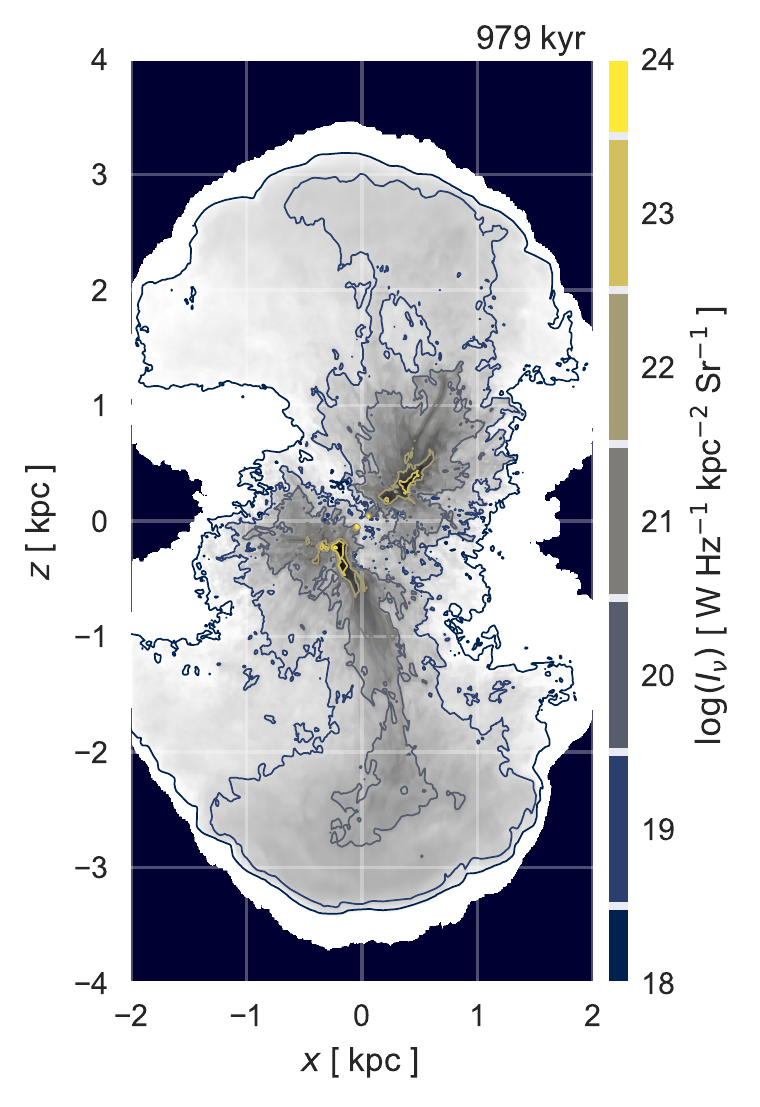}
    \caption{{\bf Left panel:} Cartoon  of the possible scenario explaining the observed properties of \target and the host galaxy NGC 1167. The host galaxy is viewed edge-on. {\bf Right panel:} Synthetic radio (1.4 GHz) surface brightness map from the simulation of the jet entering an edge-on gas disc at an angle of 45\pp{\circ}. The regions in yellow contours indicate the bright radio jets, while the blue contours are the diffuse radio emission produced by the plasma percolating through the gas disc. The image has been obtained at an arbitrary instance in the simulation for qualitative comparison. The remarkable similarity should be noted between the radio jets in this map (yellow contours) and the morphology of B2 0258+35 (Figure 1 inset) in the absence of any fine-tuning. This further supports the proposed strong jet--ISM interaction in \targets.}
    \label{fig:cartoon}
\end{figure*}

\subsection{Impact of the host galaxy on the radio continuum}\label{impact_cont}

As discussed in the earlier sections, the radio jets in \target appear to be expanding into a gas disc, injecting
a significant amount of turbulence into the disc. 
If this is the case, numerical simulations (e.g. Mukherjee et al. 2018a) predict that the ISM strongly influences the overall morphology of the radio continuum.
A strong jet--ISM interaction, in addition to generating the
shocks described above, would also launch subrelativistic winds
perpendicular to the plane of the galaxy. These winds resemble
the wide-angle outflows seen associated with AGN winds or
nuclear starburst activities (see Figure 13 in Mukherjee et al.
2018b), but are driven by the pressure of
escaping jet plasma. Buoyancy further contributes to the propagation
and expansion of the bubbles to large radii. The parameters of the jet-disc simulations 
presented in Mukherjee et al. 2018a are not fine-tuned to be an accurate representation of 
B2~0258+35. Nevertheless, simulation D in that paper, representing $10^{45} \> \rm erg \> s^{-1}$ 
jets propagating at $45^\circ$ into an inhomogeneous disc does reproduce several of the features of B2~0258+35. 

In order to facilitate the comparison, we generate a synthetic radio surface brightness image from simulation D, using the approach described in \citet{Bicknell18}; this approach was also used to simulate the jet--ISM
interaction in IC 5063 \citep{Mukherjee18b}. The main aim
of this exercise is to illustrate the effect of ISM on the radio continuum
and investigate whether the low surface brightness radio
lobes can originate from such jet--ISM interactions. The synthetic
radio image at 1.4 GHz was obtained at a time of $9.8 \times 10^5 \> \rm yr$, chosen so that the extent of the radio jets matches the observed size of the radio jets in B2 0258+ 35. The radio structure thus obtained is a result of radio plasma expanding
into a turbulent, inhomogeneous disc of gas and is shown in the right panel of Figure \ref{fig:cartoon}.

There is a remarkable similarity between the morphology of the radio jets obtained
from the simulations (see the yellow contours in Figure \ref{fig:cartoon}, left panel) and
B2~0258+ 35. In the synthetic image, the northern jet is clearly disrupted by its interaction with the ISM, but maintains a generally straight path; the southern jet has a sharp bend, which is the result of its path being obstructed by a dense cloud. There is also a hot spot in the jet approximately halfway between the core and the deflection point. The difference in the jet 
morphologies is the result of the fractal, log-normal distribution of density in the ISM, 
leading to an approximately  straight path in one jet, but to a substantial obstruction in the 
other. There is, of course, a difference between the observation and the synthetic image. In the source, the southern jet is deflected to the east; in the simulation, the southern jet is deflected to the west. This feature may be the result of the random distribution of dense clumps or the gravitational field of the galaxy and requires attention in more detailed simulations. Nevertheless, we conclude that the asymmetry
in the radio continuum is a result of a strong jet interaction with an inhomogeneous disc.

As noted earlier, this interaction also results in kiloparsec-scale
bubbles whose overall direction is different from the jets and is
determined by buoyancy of the jet plasma in the gravitational
field of the galaxy. However, the kiloparsec-scale simulations do not follow the evolution
of the jet lobes and the bubble beyond 4 kpc, so that a comparison
with the large, low surface brightness radio structures on 100
kpc scales observed in B2 0258+ 35 (see Fig. 1) is not possible at this time.
Moreover, as can be seen in Figure 10, the subrelativistic
winds have only spread out to a diameter of 6 kpc in this duration,
while the radio lobes in question are 240 kpc in size.
Hence, the faint radio structures produced by the simulations
with the present jet, galaxy, and ISM parameters do not explain
the large radio structures. As shown in \citet{Brienza18},
these lobes do not show spectral steepening, and it has 
been suggested that they represent a previous episode of activity that
switched off no more than a few tens of Myr ago, or that they are the
result of large-scale jets that have been temporarily disrupted
and smothered, but that are still fuelled, albeit at a low rate,
by the nuclear engine \citep[perhaps similar to what is happening in Centaurus A; e.g.][]{McKinley18, Morganti99}. In this scenario, the kiloparsec-scale bubbles in the model could be the connecting structure keeping the fuelling channel open. This is schematically represented in Figure \ref{fig:cartoon}, left panel.

%%%%%%%%%%%%%%%%%%%%%%%%%%%%%%%%%%%%%

\section{Summary}

We have presented high-resolution VLA-A array and EVN \HI absorption observations of B2 0258+35, a low-power radio AGN in a gas-rich host galaxy NGC 1167, which also hosts large (240 kpc) radio lobes. With our VLA observations, the continuum source and the absorption, is slightly resolved. We detect absorption all against the source, and the profile also exhibits a slight asymmetry towards the blueshifted velocities. At VLBI scales most of the continuum is resolved out, and we detect only \apx 10\pc of the total continuum seen with the VLA. We also detect a weak absorption feature in the integrated VLBI spectrum of the bright knot seen along the southern jet and the estimated \HI column density is in agreement with that estimated from the VLA observations. This implies a uniform screen of gas in front of the background radio source. 

We modelled the \HI 21cm absorption and find that the kinematics of the absorbing gas should be different from the large \HI disc. CO(1-0) emission was also  detected from the host galaxy NGC 1167 and the emission profile agrees well with the integrated \HI absorption profile suggesting that CO and \HI are co-spatial and are being affected by the radio jets to a similar extent. The line ratio CO(2-1)/CO(1-0) $\approx$ 1 further supports this hypothesis. We conclude that the radio source is interacting with the ambient gas, causing the observed turbulence, which has also affected the morphology of the radio source. Assuming the kinematics of molecular gas to be entirely chaotic, we find that the upper limit to the turbulent kinetic energy in the medium is \apx 10$^{55}$ erg. We also find that the radio jets, in this case, are capable of depositing this amount of turbulence in the medium within 0.4 Myr, the estimated age of the radio source. The optical line ratios ([O III]/H$\beta$, [NII]/H$\alpha$ and [S II]/H$\alpha$) in the regions a few kiloparsecs  from the radio nucleus suggest that the emission there arises from shocks caused by the jet--ISM interaction. All these results support the fact that low-luminosity radio AGNs are capable of affecting the ISM significantly and hence contribute to negative feedback in their host galaxies.

We further compare our results with the predictions from numerical simulations and conclude that the interactions of the jets with an inhomogeneous disc are likely to be the cause of the widespread shocks, quenched star formation,  and low surface brightness radio bubbles resulting from plasma leaking out in the direction perpendicular to the plane of the galaxy,  possibly over multiple episodes
of AGN activity. High-resolution mapping of ionised and molecular gas and targeted simulations can shed further light on the jet--ISM interaction in B2 0258+35.

\begin{acknowledgements}
We would like to thank the referee for the useful comments. We also thank Bjorn Emonts for the help with the optical spectra of \targets, and Ewan O'Sullivan and Fran\c{c}oise Combes for providing their CO emission spectrum. SM would like to thank Aparajitha Ramesh, Anqi Li, and Francesco Santoro for the useful discussions. Part of the research leading to these results has received funding from the European Research Council under the European Union's Seventh Framework Programme (FP/2007-2013) / ERC Advanced Grant RADIOLIFE-320745.
\end{acknowledgements}

   \bibliographystyle{aa} % style aa.bst
   \bibliography{ref} % your references Yourfile.bib

\end{document}